# From Text to Self: Users' Perception of AIMC Tools on Interpersonal Communication and Self

Yue Fu
University of Washington
Seattle, Washington, US
chrisfu@uw.edu

Sami Foell
University of Washington
Seattle, Washington, US
sfoell@uw.edu

Xuhai Xu
Massachusetts Institute of Technology
Cambridge, Massachusetts, US
xoxu@mit.edu

Alexis Hiniker
University of Washington
Seattle, Washington, US
alexisr@uw.edu

## ABSTRACT

In the rapidly evolving landscape of AI-mediated communication (AIMC), tools powered by Large Language Models (LLMs) are becoming integral to interpersonal communication. Employing a mixed-methods approach, we conducted a one-week diary and interview study to explore users' perceptions of these tools' ability to: 1) support interpersonal communication in the short-term, and 2) lead to potential long-term effects. Our findings indicate that participants view AIMC support favorably, citing benefits such as increased communication confidence, finding precise language to express their thoughts, and navigating linguistic and cultural barriers. However, our findings also show current limitations of AIMC tools, including verbosity, unnatural responses, and excessive emotional intensity. These shortcomings are further exacerbated by user concerns about inauthenticity and potential overreliance on the technology. We identify four key communication spaces delineated by communication stakes (high or low) and relationship dynamics (formal or informal) that differentially predict users' attitudes toward AIMC tools. Specifically, participants report that these tools are more suitable for communicating in formal relationships than informal ones and more beneficial in high-stakes than low-stakes communication.

## 1 INTRODUCTION

AI-mediated communication (AIMC) tools are rapidly gaining popularity following the development of ChatGPT and other technologies supported by Large Language Models (LLM). Hancock et al. define AIMC as interpersonal communication that is not simply transmitted by technology, but modified, augmented, or even generated by a computational agent to achieve communication goals [14]. These tools can support a variety of writing tasks, such as: idea generation, shaping tone and voice, managing multilingual translations, proofreading, and editing. Hundreds of AIMC tools, supported by LLM models from OpenAI, Anthropic, and others, have emerged this year (2023). Mirroring the rapid advancement of AIMC, user adoption of these technologies is also escalating swiftly. For instance, Grammarly, a cloud-based typing assistant company, released an AI-driven writing assistant in beta version in April of 2023 [40], and by the end of May, they reported a user base exceeding 30 million users worldwide [2]. AIMC tools are increasingly influential, with increasing evidence that AI can alter the norms of human communication [29]. For instance, Jakesch et al. showed that the use of an opinionated language model can influence participants' writing and shift their opinions on social issues [20].

Interpersonal communication serves a range of functions, and people's communication patterns can have profound implications for almost all aspects of their lives. For example, as social beings, we rely on interactions with others to satisfy essential needs for inclusion, affection, and companionship [31]. The way people speak to others can determine whether and how they are able to resolve conflicts [4], acquire information [42], build social capital [7], or make persuasive arguments [10]. Thus, the increasing adoption of AIMC tools for interpersonal communication has the potential to significantly impact both individuals and society. Prior work has reported the potential for AIMC tools to have both positive effects (such as improved communication speed and positivity [18]) and negative ones (including eroded trust [24] and adverse perceptions of others [18, 48]).

Despite growing research interest in AIMC tools, prior work has not yet examined how people are using these tools to express themselves in communication with other people or how people feel about this prospect. Here, we investigate this space, asking specifically:

- **RQ1:** What publicly available AIMC tools offer to support users' interpersonal communication?
- **RQ2:** How do users feel about their experiences using these tools?
- **RQ3:** In what contexts (if any) do users find these tools acceptable for interpersonal communication?
- **RQ4:** How (if at all) do users think the long-term use of these tools will shape their communication?

To investigate these research questions, we employed a mixed-methods approach. We conducted a one-week diary study with 15 participants who we asked to use AIMC tools of their choosing to modify at least three of their online interpersonal communications each day. Participants completed 3-5 survey entries during the diary study documenting their experiences and perspectives. We supplemented this with semi-structured interviews before and after the diary study to capture users' attitudes towards leveraging AIMC tools for interpersonal communication, their acceptance, their beliefs about potential impacts of long-term use, and their design suggestions.

Participants reported an overall positive experience using AIMC tools for interpersonal communication. They found that the tools increased their confidence in their communication, and at times, helped them voice their thoughts and feelings with greater precision. They described leveraging AIMC tools to brainstorm communication ideas, refine their messages, and interpret the emotions behind the messages they received from others. However, participants also reported common shortcomings of these tools, which frequently bloated their messages with tangential and overly emotional content. These patterns made the tools more useful in some contexts than others; participants found them more appropriate for communicating in formal relationships than in informal ones, and they found the support more helpful in high-stakes contexts where they benefited from coaching and content ideas from the tool. This work contributes empirical data about users' experiences with current AIMC tools and outlines design guidance for developing future ones.

## 2 RELATED WORK
### 2.1 The Importance of Interpersonal Communication

Interpersonal communication is the backbone of human relationships, influencing both personal and professional interactions. As Knapp [22] and Vangelisti [47] articulate, every message between two people carries two layers of information: the content level and the relationship level. Studies have shown that effective interpersonal communication enhances relationship satisfaction and contributes to people's wellbeing [12, 22, 49]. Interpersonal communication shapes the quality of personal relationships, serving as the main channel for emotional expression and personal disclosure [5, 12, 22]. Similarly, effective communication is crucial to teamwork and collaboration in professional relationships. Further, it enhances project productivity, increases job satisfaction, and reduces team conflicts [32, 37].

People begin to develop interpersonal communication skills from infancy, and learning a first language is a social process in which caregivers scaffold infants in engaging with others and expressing themselves [30]. Throughout childhood, people acquire increasingly sophisticated interpersonal communication skills, for example, learning to negotiate, hedge, or persuade [15, 50]. People continue to refine their interpersonal communication through both formal and informal learning experiences, with school-based curricula teaching, for example, expository communication [28] and socioemotional skills [19] and peers influencing communication habits [52].

### 2.2 Computer-Mediated Communication (CMC) and AI-Mediated Communication (AIMC)

People's interpersonal communication increasingly intersects with their use of digital systems. In English-speaking countries, more than 75% of adults under the age of 40 communicate more by text than they do in-person [25]. The sociotechnical affordances and culture of the system that a user engages with shape both the user's communication choices and the effect of that communication on their relationships with the people they communicate with. For example, the option to use emojis [39], the potential for threaded replies [41], the practice of using SMS abbreviations [33], and the presence of scaffolding encouraging users to reflect before writing [27] all influence the way people choose to speak to one another. Thus, understanding the impact of ICT design choices on people's communication practices is of great importance.

Walther's seminal work on Computer-Mediated Communication (CMC) introduced the concept of "*Hyperpersonal Communication*," which refers to CMC's ability to facilitate more selective and intentional self-presentation than face-to-face communication [46]. As AI technology becomes more and more available, "*Hyperpersonal Communication*" has been extended and amplified by AI-mediated communication (AIMC). Hancock et al. define AIMC as "*mediated communication between people in which a computational agent operates on behalf of a communicator by modifying, augmenting, or generating messages to accomplish communication or interpersonal goals*" [14]. Hancock et al. also offer a research agenda that calls for empirical research into the design of AIMC systems, including their impact on language, self-presentation, and self-perception. They suggest a need for work that examines both short-term and long-term impact on individuals [14].

### 2.3 Examining AIMC Systems

Early studies on how AIMC tools are used in interactions between humans, such as the work by Hohenstein and Jung [16], focused on the discrepancies between AI-assisted suggestions and actual conversational content. The study highlighted the need for improvements in AI-assisted messaging apps to make them more aligned with human communication. Recent works, including those by Liu et al. [24] and Hohenstein & Jung [17], have centered on the theme of trust and agency. The former found that people's trust in email communication decreases when they perceive the sender to be an AI agent rather than a human. The latter shows that the presence of AI-generated smart replies serve to increase perceived trust between human communicators, making AI a "*moral crumple zone*" that can absorb blame when conversations go awry.

In related research, Jakesch et al. [21] researched the impact of AIMC on self-presentation and social perception. Their findings suggested a "*Replicant Effect,*" where participants mistrust people who they believe have put up online profiles that were partially generated by AI. Poddar et al. also examine how using AIMC tools affects self-presentation, finding that AIMC may change the topics users talk about when introducing themselves to others [38]. Moreover, through an experimental design featuring suggested text responses (Google's smart replies), Mieczkowski et al. [29] find evidence of a positivity bias in AI-generated language and also note that AI-generated language may compromise aspects of interpersonal perception (like social attraction).

Collectively, these works show the importance of interpersonal communication and the complex ways in which people acquire and refine their interpersonal communication skills over their lifetime. Prior work further shows the increasingly influential role that digital systems play in mediating people's interpersonal communication and the potential for these tools to not only facilitate communication but to alter it. As AIMC tools become widespread, it is essential to examine the ways in which they alter individual and societal communication patterns. We build on this foundation and contribute to this space by examining users' experiences leveraging publicly available AIMC tools to craft their interpersonal communication, their perceptions of these tools' ability, their beliefs about the tools' impact on their communication habits, and their predictions about the long-term influence such tools might have.

## 3 METHOD

To investigate people's perspectives on leveraging AIMC tools to shape their short-term and long-term communication, we conducted a one-week diary and interview study. The study consisted of an initial semi-structured interview (N=15; one participant dropped out after initial interview), a week-long diary study (N=14), and a concluding interview that solicited design feedback (N=14). We piloted the study with two participants from the academic institution of the first author and included this pilot data in the paper (pilotA and pilotB). Including pilot participants, we report on complete data from N=16 participants, plus data from one additional initial interview.

### 3.1 Participants

We recruited 15 participants through professional and academic Slack channels and email lists, where we advertised a study on the impact of AI-powered writing tools on communication. An initial screening survey asked participants about their familiarity with AIMC tools. To qualify, participants must have had some experience using AIMC tools, such as ChatGPT or GrammarlyGO. We ensured a diverse range of users regarding experience in using these tools for interpersonal communication, from those who use AIMC tools multiple times daily to those who only use them monthly. The screening survey also asked about demographic data (see Appendix A for all participant demographics). All but one participant completed both the pre- and post-interview as well as the diary study. Each participant who completed all components of the study received a US$100 Amazon gift card as a thank-you for their participation. The participant who dropped out after the initial interview received a US$35 Amazon gift card.

### 3.2 Procedure and Materials

*Initial Interview.* We conducted an initial semi-structured online interview with each study participant. We asked about participants' current experiences with AIMC tools, how they feel about the use of AIMC tools, and their thoughts about the tools' potential influence on the participant. During the interviews, we asked participants about the frequency of their AIMC tool use, their specific objectives for using such tools, and which tools they regularly interact with. Special attention was given to the "tone" functionalities offered by various AIMC tools. We also explored participants' acceptance levels toward using AIMC tools in their communication. Next, we asked whether they view these tools as having the potential to affect their communication, and if so, how they feel about this influence. To conclude the initial interview, we introduced the upcoming diary study, outlining the tasks participants would be required to perform over the next week. Each initial interview was designed to last 45-60 minutes. All interviews were recorded for both video and audio and subsequently transcribed for analysis.

*Diary Survey.* Each participant then completed a one-week diary study. We instructed them to log a minimum of three diary entries over the course of the week. For each entry, the participant shared an example of a message or other text-based online interpersonal communication written by the participant and modified using AIMC tools before sending to the participant's target recipient. Participants were asked to share genuine communications. Participants submitted each diary entry through a Qualtrics survey form[1]. All participants received the same compensation regardless of how many diary entries they submitted.

*Training Participants in Using an AIMC Tool for the Dairy Study.* At the end of the initial interview, we gave all participants a training session on using an AIMC tool for the diary study. Specifically, we displayed a visual featuring various prompts, such as "*make it friendly,*" "*make it assertive,*" and "*express uncertainty,*" inspired by the tool GrammarlyGO. Participants had the option to choose between 1 to 4 prompts from the visual or create their own that aligned with their specific communication goals. Although participants were encouraged to use the selected prompts, we told them that that they had autonomy to adapt their prompts based on the specific communication context. They were also free to use any AIMC tools of their preference. To provide a practical example, we used ChatGPT (ChatGPT 3.5) [2] to demonstrate how to modify a sample message and record the data in a Qualtrics survey. We asked participants to input:

(1) **Recipient:** A friend, teacher, coworker, parent, etc.
(2) **Context:** Location, time, communication context, and goals.
(3) **Original Message:** Participants' originally typed messages.
(4) **User Prompts:** Participants recorded the first, second (optional), and last prompts (optional) given to the AI writing tools.

---
[1] Qualtrics Service
[2] OpenAI ChatGPT

(5) **AIMC Output:** Participants recorded the first, second (optional), and last (optional) AIMC output.
(6) **Final Message:** The version of the message sent by the participant after their optional editing.
(7) **Reflections:** Participants responded to prompts such as,"*Did the AI output align with your intended communication goals?*," "*How do you feel about the sent message as compared to the original?*," and "*Does the sent message authentically convey your intended tone?*"
(8) **Satisfaction:** Participants were asked to rate their experience using AI on a scale from 1 (dissatisfied) to 10 (very satisfied).
(9) **Feelings:** Participants recorded their feeling after sending the message on a scale from 1 (feeling not good) to 10 (feeling great).
(10) **Other Comments:** Participants were given the opportunity to include additional comments to provide more context or description about a specific instance.

*Exit Interview.* Upon completion of the one-week diary study, participants were invited to have a follow-up interview with the researchers. The exit interview first asked about their overall experience of interacting with AIMC tools, following participants' lead. We presented them with a specific survey entry from their diary study entries that we found intriguing to stimulate further reflection. We invited them to share their likes and dislikes and comment on the experience. We asked participants to explain their process of using AIMC tools during the message-modification process. Then, we asked about their perception and attitudes about using AIMC tools to shape their communication. Additionally, we explored whether they felt their use of AIMC tools influenced their self-perception or how they present themselves to others.

Finally, we concluded the interview by asking the participant to speculate about both the short-term and long-term potential and drawbacks of using AIMC tools. We presented several prototypes to elicit design feedback and design ideas (see Figure 1 and Figure 6).

The first prototype presented an AI writing tool's setup page to participants, offering features like understanding their current writing style and tone, and allowed them to select from a range of styles and tones for enhancement based on their existing style. Users could also define their target audience and choose from various AI support roles, such as a "writing optimizer," "grammar checker," "communication coach," or "explorer," each accompanied by descriptions of their distinct functions (see Figure 6). The second prototype introduced a feature enabling users to select a renowned figure whose communication style they could emulate. The feature presented analysis of each role model's style and tone, informing users that they could tailor these styles to their preferences. The third prototype demonstrated an AIMC tool's in-the-moment writing assistance, informing participants that it could analyze their email writing to detect tone and make recommendations. Additionally, it enabled the user to input contextual details, such as the conversation's goals and the relationship between the sender and receiver (see Figure 1). We also showed specific features to participants, including functions to adjust the AIMC tool's output length and the capability to incorporate emojis.

### 3.3 Ethical Considerations of Data Collection

During recruitment, we provided participants with comprehensive details about the study's design, duration, and compensation. We recognize that our goal of collecting data on interpersonal communication could involve sensitive communication topics. We informed participants that they had complete freedom to choose what to report of their communication. Although most participants used placeholders for names and other identifiable details, we observed several instances where names and locations were included in their messages. To address this, we anonymized all daily survey entries upon collection and stored the data on a secure server. This study was reviewed by our institutional review board (IRB) and deemed exempt.

### 3.4 Data Analysis

We collected two types of data from each participant: 1) daily survey data, 2) interview data from two interviews. To differentiate these data sets, we employed a labeling system. Daily survey entry data received a "-s" suffix added to the participant ID (e.g., "P14-s"), whereas for the initial interview, we used the participant ID without any suffix. Exit interview data were labeled with an "-e" suffix (e.g., "P14-e").

*Interview Data Analysis.* We coded both interview transcripts and open-ended questions from daily survey entries for emergent themes using a thematic analysis approach [8]. Each researcher initially reviewed the diary entries and interview transcripts independently. Subsequent team discussions clarified and refined these initial codes. During these discussions, we shared potential codes and code descriptions, examined example quotes collaboratively, examined counter-examples, compared code categories against one another, and refined the boundaries and definitions of each code. We used Delve[3], a collaborative qualitative analysis software program, to assist in the coding process. We used the Delve interface to asynchronously compare each researcher's codes, which we then organized and condensed into overarching themes collaboratively as a team. After multiple rounds of coding and refinement, one researcher revisited all data to extract representative quotes for each theme.

*Diary Entry Data Analysis.* We collected a total of 274 diary entries from 14 participants ($m$ = 2.6=18, $sd$ = 2.6=6, per participant). One participant, P14, misunderstood the instructions regarding the final message and AIMC output, leading us to exclude their diary data from further analysis. Additional entries lacking a final message were also omitted. After these exclusions, 226 entries remained for detailed analysis.

For the diary data, each researcher first independently examined a selection of survey entries. As we analyzed and coded the interview data, we noted a consistent pattern during the interview: participants' acceptance and experience of using AIMC tools for interpersonal communication hinged on the stakes of communication and nature of the relationship involved. We discussed and agreed to code the diary survey entry data based on these two key attributes discovered from the interview data: communication stakes (high or low) and the relationship dynamics (formal or informal). We provide definitions and examples of these codes in Table 1 and Table 2, and we coded each diary entry considering its metadata (e.g. recipient, context) and its communication content. One researcher subsequently coded all 226 entries based on the code definition. To evaluate interrater reliability, another researcher independently coded approximately 10% of the sample (23 entries). We identified only one discrepancy in a single attribute (communication stakes) between the two researchers (Cohen's $k$ = 0.91).

---
[3]Delve

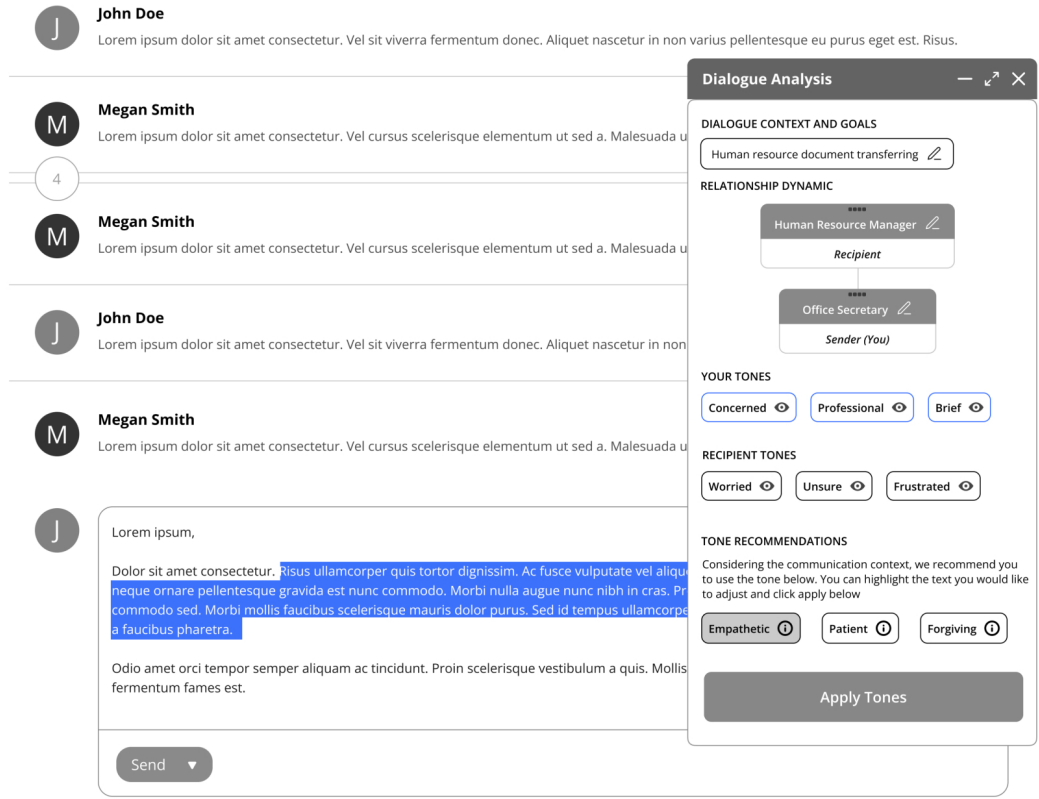

Figure 1: One of the design prototypes (prototype three) we showed to participants. It depicts an in-the-moment AIMC tool that analyzes participants' email communication and detects and suggests potential tones to the participants.

## 4 RESULTS

Over the course of one week, participants submitted a total of 226 high-quality diary entries. Overall, participants reported a generally positive experience using AIMC tools. Mean satisfaction score was 7.1 ($sd$ = 2.6) on a scale from 1–10 (see Figure 2). Notably, satisfaction scores increased after the first day, suggesting users became more adept with AIMC tools after an initial learning period (first day: $mean$ = 5.7, $sd$ = 3.1; remaining days: $mean$ = 7.36, $sd$ = 2.41; $t$ = −3.1, $p$ < 0.005).

Users' generally positive attitudes were also reflected in their answers to open-ended diary questions. For example, one participant said AIMC "*accurately edited the message and generated a response the way I wanted it to be*" (P3-s), and another one mentioned "*I feel comprehended...It worked well for me*" (P15-s). Some participants who had not previously used AIMC tools for interpersonal communication changed their attitude over the course of interacting with the tool for one week. For example, participants said things like, "*This is the first time I used ChatGPT for this purpose and I was very impressed...it was fun to be able to get poetic lines from the AI*" (P2-s).

Despite their overall positive experience, participants also reported common shortcomings of the AIMC tools they used, and they described contexts in which they found these tools unnecessary or even harmful. They also raised concerns about the potential long-term negative consequences of adopting these tools. Here, we describe these nuances and report on the patterns in participants' differentiation between positive and negative aspects of using AIMC tools for interpersonal communication.

### 4.1 User-Perceived Benefits of Using AIMC Tools for Interpersonal Communication

***Increasing Users' Confidence in Their Communication.*** Participants reported experiencing reduced anxiety, increased confidence, and a heightened willingness to communicate as a result of the support offered by AIMC tools. They mentioned that AIMC tools offer "*peace of mind*" by ensuring grammatical correctness (pilotB) and "*offload a lot of [over-]thinking*" (P9). Participants reported that this peace of mind made them more willing to communicate, saying things like, "*Even in an emotionally distressing state of mind, I was able to communicate...Without ChatGPT, I would probably do that like two days later*" (P2). Others mentioned that before using AIMC tools they "*used to overthink these things a lot...I would just delay drafting emails or responding to emails*" (P9). Participants said that having the safety net of AIMC support reduces procrastination, enabling them "*to do it much faster*" (P2) and to stop "*putting off replying*" (P8-s). Participants also explained that the confidence they gained from using AIMC tools spilled over into other experiences expressing themselves. They said that AIMC tools have "*grown my confidence as a professional writer or someone aspiring to advance in my career*" (P10). This was particularly true for those for whom English is a second language; these participants explained that they had come to feel "*a lot more confident in the emails or whatever writing I send out to people, especially to a large group of people*" (P13).

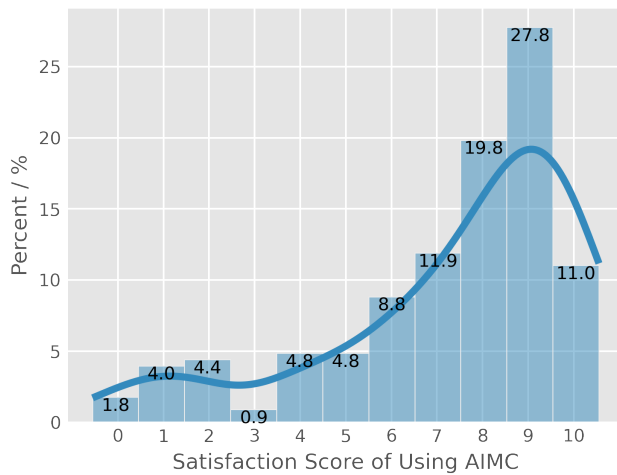

**Figure 2: Distribution of participant satisfaction in diary entries.**

***Helping Users Find the Right Words.*** In the best cases, participants reported that AIMC tools helped them find the perfect words to give voice to their ideas and emotions. As one participant mentioned, "*I think the best thing about ChatGPT is just being able to express yourself in ways that you would optimally express yourself*" (P1). Similarly, participant P11 explained: "*Sometimes I find it hard to find words for certain things. Also sometimes I feel the way I communicate or the way I write is very one dimensional...sometimes I really like certain sentences [that] ChatGPT gives. I wouldn't have thought of those sentences*" (P11). Another participant stated, "*It'd give me a phrase or a way to word something that I didn't think about before. And I'd be like, 'Oh, I totally forgot that phrase existed. This is perfect'*" (P10). This iteration with AIMC tools helped users to refine their thoughts over time and articulate vague or blurry ideas. Participant P12 mentioned that ChatGPT could "*tidy up ideas*" and make them

"*easier to process,*" while participant P1 found that the AIMC tool was instrumental in "*forming the correct words and the correct ideas.*" P2 described AIMC tools providing the right words when they were just out of reach, saying, "*I find it extremely useful being able to find the right words, especially because I usually think I know the words, they're just not off the top of my head, and the ChatGPT can usually help me with that.*" In these and other instances, participants said that AIMC tools helped them to express themselves more precisely and authentically.

***Supporting Cross-Cultural Communication.*** Participants noted that AI can help individuals navigate the complexities of language and culture and bridge communication gaps. One participant stated, "*People use those [AIMC tools] to help them to sound in a native way or help them to express themselves correctly to other people, especially with different cultural backgrounds*" (P6).

Another participant described composing a culturally appropriate birthday message, saying, "*I just crafted the messages in ChatGPT [telling it] I'm a Muslim I want to give it a very Muslim touch to a message...It actually gives you information like that even it gives you the information from the Holy Scripts*" (P15). In these instances, participants reported that AIMC tools can play a important role in making communication more culturally appropriate, thereby improving understanding and sensitivity. However, not all participants agreed that AIMC tools are currently sophisticated enough to perform this function well. For example, participant P10 explained, "*I'm personally from New Jersey, which is very close to New York and the Jersey shore. We pretty much have our own language...So the fact that everyone's so different and from different communities and as they speak and write, they are really hard for AI.*" Despite this skepticism, many participants saw potential for current and future AIMC tools to enhance people's ability to communicate effectively across cultures.

***Improving Users' Grammar and Vocabulary.*** Participants reported that using AIMC tools broadened their vocabulary and understanding of language. Participant P9 observed, "*Most of the time you are short on words, especially if you don't really have an extensive vocabulary. But when you're using AI writing tools, it gives you a sense of how you can structure your sentences and what are all the words that you can commonly use. I think it extends your vocabulary*" (P9). Similarly, P7 stated, "*It'll definitely enhance my vocabulary a lot. It influences my choice of words when I speak even in the real world*" (P7). Participants also said that AIMC tools led them to actively learn to express themselves more clearly, saying things like, "*I would go ahead and find the meanings of those words [generated by AIMC]. And that ended up in me learning the meaning of a lot of words and using it in my day-to-day communication as well*" (P7-e). Another participant agreed, stating, "*I've learned new words by using ChatGPT, and I've integrated those as part of my vocabulary*" (P12-e). Other participants noticed improvement in their phrasing and syntax, saying things like, "*I think it did really help me to improve my grammar...I noticed that because of the repetitive usage of ChatGPT, it seems like my language improved a lot*" (P3).

***Helping Users Brainstorm.*** Participants also said that they used AIMC tools to help them generate new content and brainstorm ideas for communicating with others. Participants described this process saying things like, "*The AI created something in the case where I didn't really know where to start. It was helpful, but then I think it still required some iterations on my side*" (P12-e). Users said that AIMC tools helped them think more expansively, saying things like, "*Sometimes AI adds stuff, and I feel like that would push you to think, 'oh yeah, this is also a great question to ask, maybe I could add this to the message as well'*" (P3-e). In these instances, participants explained that they would let AIMC "*brainstorm certain things*" and then they would "*edit on top of that*" (P7-e). P14 similarly described that AIMC would "*help me come up with new ideas or a new way that I could potentially respond to a specific message*" (P14).

***Supporting Emotionally Aware Communication.*** Separately, participants used AIMC tools to help them build awareness of others' emotions and to workshop the emotional impact of their own messages. One participant stated, "*I took that text to…ChatGPT and I was like, 'Hey, if I send this to them, what would they think?'*" (P2). Similarly, another participant used AI to confirm the appropriateness of their language, stating that AIMC provided "*positive reinforcement*" (P12). P2 described asking an AIMC tool, "*this is what this person sent to me. What do you think it means?*" They further explained, "*once I had that comfort from ChatGPT saying that, 'oh, it's actually fine,' that then I felt actually good about like, sending that text across*" (P2). In these instances, people leaned on AIMC tools to reflect on and understand the emotions and tone they were conveying to others and, similarly, to interpret the emotions behind the messages they received from others.

## 4.2 User-Perceived Drawbacks of Using AIMC Tools in Interpersonal Communication

***Verbosity.*** Participants consistently brought up the problem of verbosity when discussing their experiences using AIMC tools for interpersonal communication. They often pointed out that the output generated by AIMC tools was longer than they had intended and bloated with tangents. For example, one participant complained, "*I wonder why the AI tool always makes the answers very long unless I prompt them to be short and concise*" (P4-e). Participant P8 corroborated this sentiment by stating, "*I think it also sometimes makes the message too superfluous…So it's like too lengthy typically*" (P8) and P11 noted, "*it generates extra content, some sentences that are not relatable*" (P11-e). Participants often edited out this extraneous content and explained, for example, "*I think I've deleted most of that and just kept what I felt was necessary*" (P7-e), underscoring the burden participants faced to simplify output from AIMC tools.

***Unnatural Output.*** Participants reported that they found the language patterns of AIMC tools to be artificial and formulaic. For instance, participant PilotB described opening phrases generated by AIMC tools by saying, "*it starts with 'I hope you're doing well' and 'I hope this email finds you well' and blah blah blah. A lot of things that normal human beings won't ever use*" (pilotB). Participant P13 explicitly observed that the generated replies seemed mechanical, saying, "*I think it sounds like something generated by AI, it sounds very templated.*" In the follow-up exit interview, P13 elaborated, "*I think Grammarly, in general, just gives me a very chatbot feeling…it just sounds very templated*" (P13-e). Similarly, participants explained that output from AIMC tools lacks "*that, uh, human touch to a message*" (P15), and "*always looks very formal and machinery kind of tone*" (P6).

***User Burden.*** When using AIMC tools, some participants reported needing to iterate on messages multiple times. The tools' failure to produce acceptable output in the initial attempts led to frustration and increased workload. One participant noted she had to "*go back and forth many times*" (P12), while another reported iterating "*five, six times…it's never a one-time thing*" (P14). They emphasized the draining nature of the iterative process, stating, "*I get a little frustrated, I've been here for a while*" (P12). Specifically, participants found these iterations burdensome because of the work to craft careful prompts. One participant mentioned, "*I feel like I have to prompt it well, so if I don't prompt it exactly what I want, they come out with really either long text or something that it was not originally what I expected*" (P4-e). Participant P9-e advised being explicit with the tool about what *not* to say, noting, "*ChatGPT ends up doing a lot more than what you ask for, so you have to be very specific with your prompts.*" For many participants, this led to the work of setting additional boundaries as they gave instructions to the tool, asking for messages that are, "*more friendly but keep it formal*" (P3-s) and "*conversational but spartan*" (P2-s). Participant P5 similarly asked one tool to "*Rewrite this in a clear, loving tone. Don't add any language that is extremely flowery*" (P5-s). Participants said that it was only with this burdensome refinement that they were able to guide AIMC tools to produce output they were satisfied with.

***Excessively Emotional Content.*** Many participants said that AIMC tools amplify emotional language, misrepresenting users' actual emotions and making their messages sound unnatural. As one participant noted, "*[AIMC tools] tend to go to the extreme side of the emotion…friends who know me will know that's not me being that extreme or expressive in emotion*" (P13-e). Similarly, another participant explained that when they request the tool use a certain tone, "*it's just too exaggerated*" (P12-e). Participants consistently said that AIMC tools would deliver the requested tone but amplify it beyond their intentions, explaining that "*if I say 'informal,' it gives me something that feels too informal…I don't know how to prompt it to gimme some, some sort of a balance*" (P11) and "*it always goes to these extremes when I want to make the email 3% more polite, it makes it so over the top.*"

***The Guilt of Ghost Writing.*** Participants expressed concerns about the authenticity and superficiality of interpersonal communication when using AIMC. PilotA, for example, argued that manually crafting a message provides a more genuine expression, stating, "*I always feel that when I'm contacting a friend or a family member, there's more sincerity if I type it myself compared to having AI generate or revise*" (pilotA). Other participants agreed, saying they want to be "*[my] authentic self to give my friends and my family my own genuine response*" (P14) and that they "*wouldn't want to change it. I would still keep my own style*" (P4-e). This led participants to feel conflicted or guilty about using AIMC tools for interpersonal communication, which participant P3 described as "*cheating.*" Another explained that using any AI as a substitute for the user's own voice would be inauthentic, regardless of the system's capabilities,. They said, "*I think it is smart enough and learning enough to know how to create different types of speech. But it still feels weird, well,* it's not authentic. So I think no matter it's level of ability, it couldn't make things more authentic for me" (P5). Some participants said that providing sufficient direction to the tool alleviated this guilt, because the AI would be "*helping me, but not replacing me*" (P12-e). Similarly, participant P9 described their approach, saying, "*I will pick up the lines that I really like from the response and then I'll add it to my message just to make it sound more personalized.*" After a week of use, one participant changed his view and explained, "*At the beginning when we first interviewed, I was like, 'I would never want to do that cause it would feel inauthentic.' And I realized like after doing it, because I was still the one crafting the text, it actually didn't feel that inauthentic to use an AI*" (P5-e). Despite these counterexamples, many participants felt that the use of AIMC tools precluded using an authentic voice.

## 4.3 Where AIMC Can (and Cannot) Support Interpersonal Communication

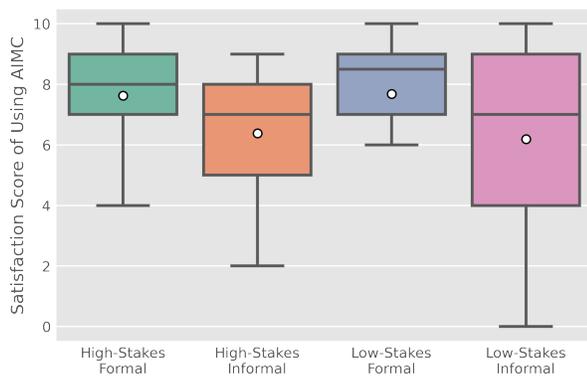

Figure 3: User satisfaction scores with different communication stakes and relationship dynamics.

After our initial analysis and discussion of the qualitative data, we found that participants' messages clustered into two categories along each of two axes: first, whether the communication stakes were *high* or *low*, and second, whether their relationship with their communication partner was *formal* or *informal* (see Tables 1 and 2). We found [N=129 (56.8%)] of messages were sent in a formal, high-stakes context, [N=29 (12.8%)] of messages were sent in an informal, high-stakes context, [16 (7.0%)] in a formal, low-stakes context, and [53 (23.3%)] in an informal, low-stakes context. We found differences in participants' satisfaction with the final AI output as a function of both communication stakes and relationship type.

For "*Communication Stakes,*" participants' satisfaction scores were higher in high-stakes communication contexts (median =8.0, mean=7.4 ± 2.4) than in low-stakes communication contexts (median=7.0, mean=6.5 ± 2.9). Similar trends were observed for "*Relationship Dynamic,*" where participants reported higher satisfaction scores when sending a message to a recipient with whom they shared a professional relationship (median=8.0, mean=7.6 ± 2.3) and lower satisfaction scores when sending a message to someone with whom they shared an intimate relationship (median=8.0, mean=6.3 ± 2.8). We ran a GLMM with "*Communication Stakes*" and "*Relationship Dynamic*" as main factors. Our results indicate the significance of "*Relationship Dynamic*" ($\chi^2(1) = 4.4, p < 0.05$), but not "*Communication Stakes*" or their interaction.

*4.3.1 Usefulness in Formal Contexts.* Participants' qualitative responses reflected some of the reasons for these distinctions. They disproportionately expressed appreciation for AIMC tools in formal contexts, saying, for example, that tone adjustments from AIMC were very useful when messaging "*a teacher or a recruiter*" (P12-e) and "*for professional communication*" (P14). Participants felt that AIMC support was particularly helpful in formal environments because of the commonly understood norms in these settings. P5 articulated that in "*professional settings, we all have an understanding that our relationships are different, and there's more purpose,*" making it possible for AIMC to do "*a good job of expressing professionalism*" (P5).

Participants reported that AIMC tools are effective in supporting formal interactions in both high- and low-stakes contexts. One participant explained that the tools are beneficial "*in the more professional context [when] there's maybe higher stakes and you want to make sure everything is properly structured and articulated*" (P7). Another described the value of using AIMC tools to message professional contacts "*when the stakes were higher*" (P12-e). Other participants described AIMC tools' effectiveness in low-stakes, formal interactions. They described the support they had received for low-stakes messages in formal contexts, saying things like, "*I did like the overall edit that it gave me, I felt like it did add more confidence and friendliness to my text*" (P5-s). Other participants said that they were "*pretty satisfied with the answer with the right emojis,*" (P4-s) to co-workers, and that the AI made them "*seem more excited and approachable, which is a good habit for sending messages on text, particularly since I mostly work with this person remotely*" (P8-s).

*4.3.2 Intrusiveness in Informal, Low-Stakes Contexts.* In informal, low-stakes contexts, participants saw little need for AIMC tools and envisioned them as more disruptive than helpful. P4-e spoke of their communication with close family, saying, "*I wouldn't want to change the way I usually speak or write stuff.*" Participant P7 emphasized the importance of spontaneity in low-stakes, informal contexts, saying, "*I type a message and I just send it. I don't think twice if it's a casual message.*" Participants P5 and P13 echoed this sentiment, mentioning that they "*didn't feel it [using AIMC tools] was necessary*" (P5) and "*didn't find it important to pay attention to all aspects*" (P13) when the communication was in a low-stakes, informal context.

The underlying reason for this preference seems to be a desire for authentic interactions, unaltered by an outside mediator. Participants mentioned "*I need to show my own identity*" (P11), and "*in my day-to-day interactions or in my genuine connections, I don't want any filter,*" highlighting the importance of genuinity for participants. Participants also explained that they do not feel the need to aim for perfect communication in this context, saying things like, "*in casual interactions, it's just more fluid, and whatever comes to your mind, it's less important to be perfectly structured*" (P7).

For two use cases, participants found some utility in using AIMC for informal, low-stakes contexts: adding appropriate emojis (P4, P14, P3), and checking mistakes during hasty communication (P9). However, participants also noted that AIMC often generated an excessive number of emojis and expressed concern that generic emoji suggestions may not be well-received (P7, P10).

Table 1: Communication spaces categorized by communication stakes (high, low).

| Communication Stakes | | |
|---|---|---|
| The level of risk and potential outcome associated with the communication between the user and a recipient. | | |
| Category | Definition | Example |
| High Stakes | Communication involves significant risks and potential outcomes. | Expressing that I'm ashamed; trying to restore trust in a friendship. |
| Low Stakes | Communication involves minimal risk and impact. | Asking parents where they are going on a trip; making casual event plans. |

Table 2: Communication spaces categorized by relationship dynamics (formal, informal).

| Relationship Dynamics | | |
|---|---|---|
| The nature and closeness of the relationship between two parties. | | |
| Category | Definition | Example |
| Formal | Governed by workplace norms, roles, and responsibilities, typically avoid personal sharing. | Messaging a potential landlord about renting an apartment; cold email to a recruiter. |
| Informal | Characterized by casual language, emotional closeness, personal sharing, and fewer boundaries. | Messaging a friend about losing a wallet. |

*4.3.3 Mixed Effects in Informal, High-Stakes Contexts.* Participants expressed diverse opinions about the use of AIMC tools in informal, high-stakes contexts. In these instances, they, for example, had online conversations with family or housemates about financial deficits, apologized to friends for inappropriate social behavior, and expressed empathy in emotionally fraught scenarios (see Figure 4). Participants explained that these high-stakes interactions required thoughtful communication, saying things like, "*If your friend is a manager at some company and you are seeking help for a referral, you cannot just write casually to him*" (P15). Participants expressed conflicting attitudes regarding AIMC's role in formal high-stakes communication. P10 expressed a clear aversion: "*I would literally hate if someone used AI to try to reject me romantically.*" On the contrary, P12 suggested that in emotionally charged situations like a falling-out between friends, AIMC could help craft a more careful response. Participant P4 also mentioned, "*If you're in some friendship difficulties and you don't know how to handle it, I think ChatGPT will be a great [tool] to give you some ideas on how to give you answers to reply*" (P4). Other participants agreed, explaining that AIMC tools could play the role of a wise friend:

> "*So previously I would just rather rely on my my closest friends, I would ask them, Hey, how do I respond to this? What do you think is a nice way to turn it down without sounding rude, but now I just might ask ChatGPT and not bug my friends that much*" (P9).

These findings suggest an opportunity for AIMC tools to support informal, high-stakes conversations in some instances, but they also demonstrate the delicate nature of these interactions, which participants said require both authenticity and thoughtfulness.

### 4.4 Users' Speculations about the Long-Term Use of AIMC

***Impact on Communication Style.*** Participants speculated about the potential for long-term changes to their communication style as a result of using AIMC tools, as they reported noticing short-term changes after just a week of use in the study. Participant P8, who had never used AIMC for interpersonal communication, noticed the tool "*has definitely changed my voice*" in certain situations after just one week of use. Participant P3 compared the process to introducing "*new habits and new things*" to a baby and observed changes in their texting style, including increased use of emojis. Another participant confirmed the tool had already influenced their writing style by saying, "*It's not a hypothesis for me anymore. It already happened or influenced my writing style. I definitely expect it will change [my] style more in the future*" (P13).

Some participants attributed these shifts to their self-reflection while using AIMC tools. Participant P12-s found it helpful to "*be critical about things I'd change or keep*" by reflecting on the output from ChatGPT. P8-s reported using AIMC tools to word a message to their mother more kindly rather than writing in their usual "*passive-aggressive*" style.

***Impact on Self-Perception and Identity.*** Most participants also predicted that long-term use of AIMC tools would influence other aspects of their self-presentation, identity, or attitudes. P14 speculated:

> "*If you're responding to a message more supportively, I think in the long run you're actually conditioning yourself somewhat to make yourself think you're a more supportive person. That definitely doesn't come, like, with a one or once or twice type of a text. But I think in the long run, if you consistently do that, then you will start to alter your self-perception in a way that mirrors*

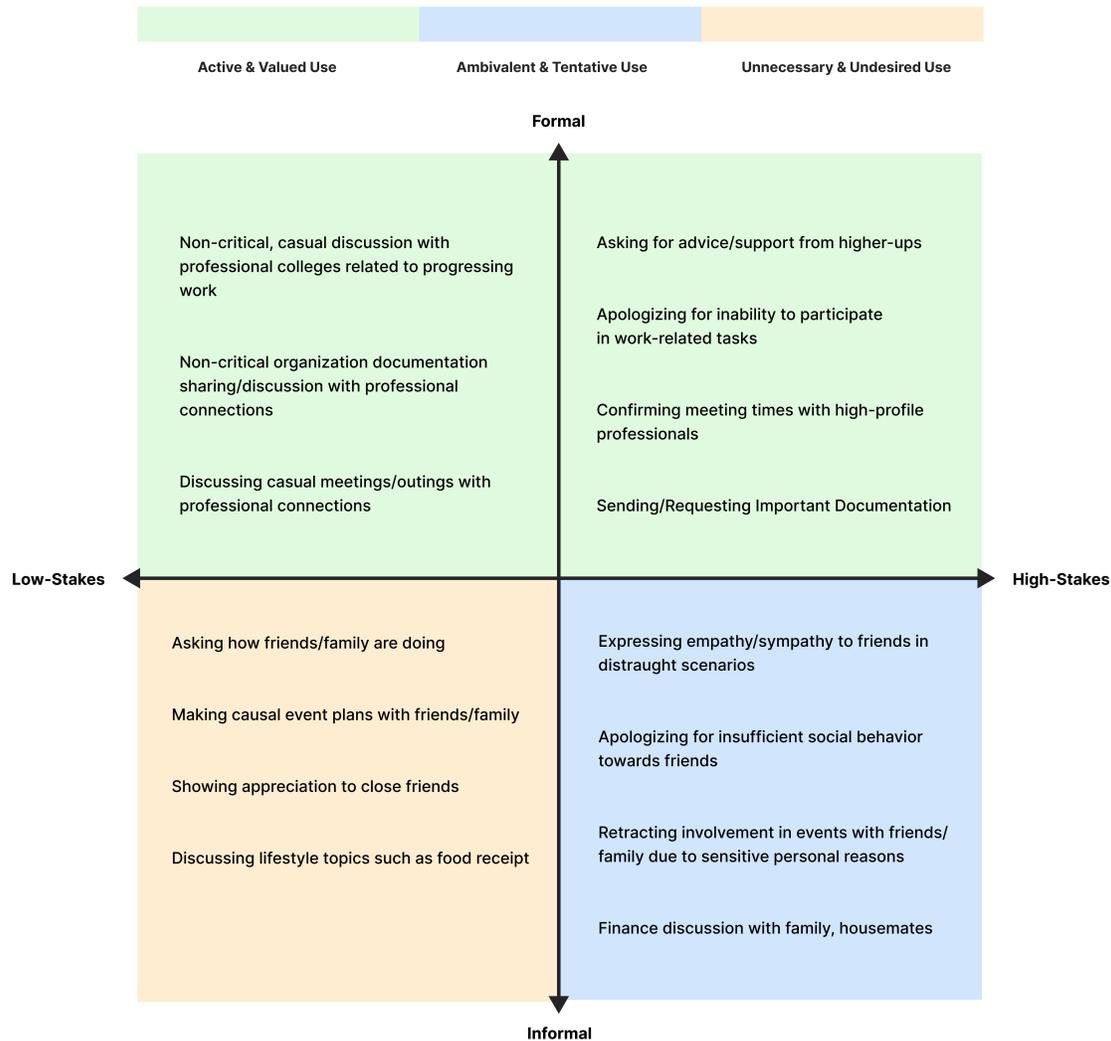

**Figure 4: Four communication spaces: formal low-stakes, formal high-stakes, informal low-stakes, and informal high-stakes.**

*the types of responses that the AI is generating for you*" (P14).

Others were less convinced that the tools would have such a dramatic effect, saying things like, "*I feel like it [AIMC] won't change the personality. I wouldn't say. I think for me, personality is something that is kind of inherent*" (P4-e).

Participants were mixed in their attitudes toward potential changes. P14 was optimistic, stating the tool "*brings out a lot of sides to you and traits in you that you perhaps hadn't honed in on previously.*" P11 felt change "*might actually be happening subconsciously. I'm using it every day. So I'm sure it, it, it will definitely impact my personality*" (P11). Some participants expressed concerns about the potential for long-term erosion of personal agency, saying things like, "*[it's] kind of like we are being controlled by their [suggested] emotional expressions*" (P6). Thus, many participants' short-term experiences with AIMC tools left them with the impression that continuing to use the tools would gradually shape how they engage with the world, leading to internalized changes. Participants saw human agency as critical and hoped for a world where these changes were driven by human judgment.

***Overreliance on AIMC Tools.*** Across participants, a common concern was the fear of becoming too dependent on AIMC, which some believed could result in intellectual stagnation. Participant P15's observation encapsulated this concern: "*I feel like it makes you dumb...if you are fully dependent on ChatGPT, it makes you dumb. You're not using your brain sometimes.*" (P15). Another participant similarly acknowledged becoming "*over-reliant*" on AI for writing tasks because, "*what I did right now is I just write whatever main message I want to convey in the email and I would just leave the rest to AI*" (P13). The concern extended to the potential for long-term influence, as participant P15 explicitly worried, "*After one month, you'll become dumb,*" while participant P9 worried about what would happen if services like ChatGPT or GrammarlyGo were suddenly no longer available.

This concern about dependency also appeared in participants worrying about transferability to the real-world face-to-face interactions. Participant P9 pointed out that while AIMC makes you sound "*professional on the text,*" it could create a disparity in how you sound in person by mentioning, "*I worry about that this sounds too good on paper, will you be able to deliver in person?*" (P9-e). Others also echoed this sentiment, worrying about a potential "*disconnect*" (P7-e) or "*difference*" (P10) between online and offline communication due to reliance on AIMC.

However, this concern about dependency coexisted with a recognition of AIMC's educational potential. Participant P2 noted that despite initial reservations, the tool had "*simultaneously been a little bit educational.*" Similarly, participant P4 worried about becoming "super dependent" on AIMC tools, but also appreciated the opportunity for learning efficient communication. This duality was echoed well by another participant who mentioned "*I'm really worried that I'm dependent on these tools, but I'm also happy that I'm learning on my way*" (P3-e).

## 4.5 Design Feedback and Suggestions from Participants

Participants expressed diverse preferences for the roles they want AIMC to adopt in their interpersonal communication and suggested multiple design features. We described their feedback and suggestions below.

*4.5.1 Global Customization and Real-Time Contextual Support.* They advocated for customization options that would allow AIMC to serve different functions, such as grammar correction or communication coaching. One participant (P12-e) took this further by suggesting creating multiple user profiles to suit different communication scenarios, like work mode or personal mode, thus advocating for global customization of AIMC personas.

In addition, real-time, in-the-moment support was a recurring theme among the participants. Participants stressed the importance of context, arguing that it is "*necessary for that whole understanding*" (P5-e). Participants suggested that AIMC could analyze the previous messages or email threads to predict possible responses during real-time communication.

Participants also highlighted the need for an intuitive and simplified user interface to minimize mental load during real-time communication. Participant P7-e appreciated quick and efficient suggestions that would allow them to "*proceed without having to go through all these [selections],*" indicating that real-time support should be straightforward to use.

*4.5.2 Preserving and Enhancing Individual Communication Style.* Several participants valued their unique communication styles and suggested AIMC should enhance but not replace their personal styles. They said things like, "*I would like to supply it [AIMC] with a bunch of my own texts, so that it could learn specific words and specific styles that I would tend to use. It's not like something completely alien to me,*" (P5-e), and "*I found that I want it to sound more like me but better in most contexts*" (P9-e).

*4.5.3 Control over Message Length and Intensity.* As mentioned in Section 4.2, verbosity is a common problem of current AIMC tools. Participants explained that they want to control the length of output. They mentioned "*the message length really mattered, which really affected what I felt*" (P7-e). Similarly, as mentioned in Section 4.2, extreme expressions from AIMC leads to low satisfaction. Participants also called for the ability to fine-tune the tone-intensity of messages.

*4.5.4 Tone and Voice Analysis.* Participants expressed a desire for tools that analyze the tone and language from the perspective of the message recipient. Participant P10 said, "*I would love to hear, like, what my tones sound like as well as the recipients'… Well, 'cause I would always like want to know how my stuff would be received.*"

*4.5.5 Tracking Progress Toward Communication Goals.* Participants also proposed a feature to track user progress toward specified communication goals. For example, participant P4-e recommended periodic reports that show improvements over time, providing users with a tangible sense of their communication growth.

*4.5.6 Explainability and Educational Features.* Participants expressed interest in understanding the mechanics of AIMC tools. For instance, P4-e said, "*I really want to know how it worked. I have this text and then how they really make it very good.*" Participant P13 appreciated the explanation function of an AIMC tool they used by saying "*And also the [AIMC tool] output will give me not only the refined email, but also how they did it.*" Participant P4-e also emphasized the importance of examples to help users grasp how different parameters function, stating, "*I think having examples will be nice. I have this text and then I use all these parameters, and then if you can give me an example of it, I kind of understand the parameters.*" P1 summed up the educational potential of these explainability features by pointing out, "*I think the biggest feature would just be users would be able to learn from using that tool or that functionality.*" These comments illustrate a consensus among participants for AIMC designs to be not just explainable but also educational, offering users insights into both the 'how' and 'why' of effective communication.

## 5 DISCUSSION

### 5.1 Design Implications for AIMC in Supporting Interpersonal Communication

Our study reveals a generally positive user experience with AIMC tools for interpersonal communication. Participants, some of whom were initially skeptical about or unfamiliar with AIMC, expressed a shift in their attitudes. They saw numerous benefits including reduced communication anxiety, enhanced willingness to communicate, improved communication quality, and support for navigating cultural and linguistic differences. Despite these positive experiences, the study participants expressed concerns about various drawbacks in current AIMC tools. Given these benefits and challenges, future AIMC designers should consider building intuitive and simple tools, scaffolding user experiences and helping users build mental models in understanding AIMC's role, for example, by providing constrained choices and comprehensible feedback to align with users' communication goals.

Across participants, we found users value their unique communication style, and expressed willingness to use AIMC if it aligned with their style. Any significant deviation, such as extreme tonal intensity or unnatural output from AIMC, led to dissatisfaction. Most were open to sharing conversational data for the purpose of training more personalized AIMC experiences. We also noted the tension between participants' concern that AIMC might represent them inauthentically by using language the participant would never use and participants' appreciation for AIMC's ability to suggest appropriate words and expressions. We recommend that designers focus on offering input options to help AIMC better capture users' nuanced intentions. For instance, an "explorer" mode could generate and present various alternative communication output options, while a "personalized" mode could align AIMC's output with the user's own communication style.

In addition, we also identified the design dichotomy between global customization and real-time support. Global customization acts as a baseline, reflecting users' general communication styles and goals, whereas real-time support allows fine-tuning to fit diverse communication contexts and offers specific communication instructions. Similar features have been launched by AIMC tools in the market since our study commenced. For instance, ChatGPT rolled out its "Custom Instruction" feature [4], and the Chrome extension Sider.ai [5] integrated options for format (email, message, twitter, etc.) and length (short, medium, long).

Along similar lines, we recommend designers consider additional parameters, such as power dynamics between communication parities, adjacency pairing [29], and recipient-group size. Future research is needed to explore: 1) whether and how LLMs can interpret these parameters for desired output, 2) how their inclusion could affect user experience, and 3) how to incorporate these design elements in both global customization and real-time support features.

## 5.2 Design Spaces of AIMC tools for Interpersonal Communication

Our interpersonal communication is shaped by relationship dynamics, including factors like familiarity, trust, emotional connection, and power differentials. These elements guide our choice of words and language styles, and how we convey information, express feelings, and manage conflicts, affecting both in-person and online exchanges. Likewise, the stakes in communication also play a crucial role. For instance, high-stakes conversations, often marked by intense emotions and varied opinions [34], require elevated metacognition and emotional control. Without adequate training and effective communication tools, individuals may default to responses driven by instinctual fight-or-flight reactions, potentially harming interpersonal relationships.

Our study, integrating interview and diary survey data, categorizes communication into four communication spaces based on these dimensions: formal high-stakes, formal low-stakes, informal high-stakes, and informal low-stakes.

---
[4]ChatGPT release note
[5]Sider.ai

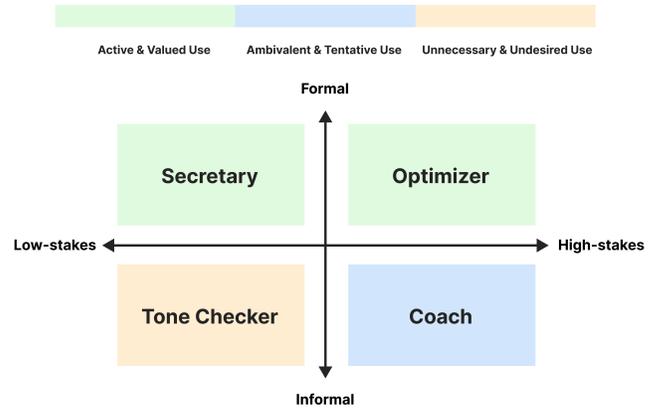

Figure 5: Different roles of AIMC in four communication spaces: secretary, optimizer, tone checker, and communication coach.

*5.2.1 Formal Communication Spaces: High-Stakes and Low-Stakes.* We find two formal communication spaces where AIMC tools excel in supporting interpersonal communication: formal high-stakes and formal low-stakes. Consistently, users reported satisfaction in leveraging AIMC tools for these scenarios, aligning with prior studies. The emergence of AIMC tools aimed at improving communication in these spaces is noteworthy. For instance, healthcare professionals found ChatGPT valuable for communicating complex and sensitive information to patients, enhancing overall patient experience [23]. Similarly, numerous resume builder tools enable users to build "professional-looking resumes in no time" [1].

Our study shows that during formal communication, users seek assistance that not only improves their writing but also aligns with their personal style. They prefer to avoid the mechanical and generic responses associated with AI. This preference highlights a disconnect in understanding. For example, though prior research indicates that AI-enhanced resumes may increase job seekers' chances of being hired [45], the widespread adoption of AIMC in formal writing raises unanswered questions about potential biases of recruiters against AI-assisted applications. These concerns may stem from issues related to trust [21, 24] and lack of personalization.

During formal, low-stakes communication, users are more sensitive to tools' ability (or inability) to align with their personal style, often through syntactic habits like using lowercase letters or emojis. In these instances, users also express a need for fast communication, suggesting a demand for AI tools that offer personalized quick responses, auto-correction, and efficient AI-mediated voice-to-text interaction.

Our study also reveals that, in many situations, there are notable subcategories within the broader categories of "formal" and "informal" relationships. This variation often depends on the nature of the relationship (e.g., a relationship with coworker vs. a boss) and organizational context (e.g., a small teams vs. a large corporate group). Current AIMC tools often lack the subtlety to navigate these nuances, typically offering generic "professional" or "formal" prompt suggestions without considering the intricate levels of formality.

Therefore, we propose that AIMC tool designers focus on: 1) building customization features tailored for various use-cases in the formal communication space, and 2) provide nuanced controls for tone and message length. To support formal, high-stakes communication, AIMC tools can function as a "*communication optimizer*," enhancing clarity and expression, while in formal low-stakes scenarios, they can serve as a "*communication secretary*," which provides functions for personalized, quick writing assistance.

*5.2.2 Informal Low-Stakes Communication Space: A Balance of Authenticity and Expediency.* Our results indicate that users generally do not perceive a need for AIMC tools in informal low-stakes communication. Writing messages with AIMC in this context is often considered inappropriate and can lead to ridicule. For instance, a prominent tennis player faced criticism from his Twitter audience for allegedly using ChatGPT to post a congratulatory note to a fellow top player [3]. Our participants expressed a desire to maintain their individuality, authenticity, and unique communication styles. However, they also acknowledged instances where quick, pre-formatted responses could be beneficial. For example, AIMC tools could offer rapid replies based on the communication context or conduct tone checks to prevent hastily sent, unintentionally uncivil messages [44]. Therefore, in this space, AIMC may function as both a "*communication expediter*" and a "*communication tone checker*."

*5.2.3 Informal High-Stakes Communication: Navigating Nuance and Sensitivity.* Informal, high-stakes communication is perhaps the most intimate and complex context for interpersonal communication. People use computer-mediated communication (CMC) for this type of communication, for example, announcing breakups through social media posts or changing their relationship status on Facebook [13]. There is a growing trend to seek online communication support, with an enormous amount of communication advice available on topics ranging from how to talk about breakups, to discussing financial issues, sending grief messages, delivering layoff notices, or responding to rebellious teens.

Our research shows that participants are open to consulting AIMC tools in this context, which they likened to the experience of consulting a friend for advice, but they resist the idea of AI crafting responses. Our results suggest that designers will best support these communication scenarios if they prioritize users' autonomy and offer multiple alternatives and neutral commentary about each. Additionally, AIMC tools can leverage the vast knowledge base of relationship science to scaffold users in better understanding their communication goals and strategies. Our results also suggest AIMC tools can bolster users' confidence and willingness to engage in this challenging communication context. Therefore, designers should focus on providing validation and educational feedback, such as detailed explanations and coaching, to support users in crafting their communication.

Finally, since informal, high-stakes communication often involves strong emotions and varied opinions [34], AIMC tools also can offer reflection and support for emotional regulation, contributing to users' well-being. In this space, AIMC can serve as a "*communication coach*."

## 5.3 Long-Term Effects of AIMC Use

We are formed by the language we use. The concept of linguistic relativity, or the Sapir-Whorf Hypothesis [26], argues that the language we speak significantly impacts our cognitive processes. Numerous studies support the notion that our choice of words reveals essential aspects of our social and psychological states [36]. These choices are often not random; patterns in language use can uncover our thoughts, emotions, and even personality traits [9, 35, 43].

There is good reason to believe by changing our language used in interpersonal communication can change how we interact with each others and influence our self-perceptions. Theories like self-perception theory suggest that individuals form attitudes based on observed behavior [6]. Similarly, the Proteus Effect posits that our behavior can be influenced by our digital self-representation [51]. And CMC has been shown to induce identity shifts [11].

Building on this, as online interpersonal communication becomes increasingly supported by AIMC tools, they could act as catalysts for change. These changes may extend beyond language styles to also influence individuals' identity, self-perception, and potentially even personality. Our exploratory study provides several indications that these shifts are already underway due to AIMC tool usage. Participants noted changes in their word choices, grammar, and language compared to their pre-AIMC communication. While participants expressed mixed feelings about long-term impacts—ranging from fear of over-dependence to optimistic anticipation—there was a general consensus that AIMC use will likely have some effect. Based on our findings, we identify several design suggestions related to the long-term effects of AIMC-use on individuals that HCI community can explore further:

- **Balancing Learning and Dependency:** Exploring how to mitigate the risk of over-dependence on AIMC while maximizing learning opportunities.
- **Supporting and Augmenting Self-Expression:** Supporting and augmenting users' personal styles and self-expression rather than replacing them with automation.
- **Including Feedback and Reflection Mechanisms:** Implementing intentional interventions with metrics and feedback to empower AIMC users to communicate more effectively and mindfully.
- **Considering Different Demographics:** Addressing different needs of different populations such as children, teenagers, mental health patients, and non-native speakers.
- **Designing for Transferable Learning:** Facilitating the transfer of learned skills from online to offline communication.

In our study, we recognize the uncertainty surrounding AIMC's long-term influence on users. Technologies often unintentionally mold human behavior and interactions. This fact emphasizes the need for in-depth research into how AIMC tools may influence people over time. The potential for AIMC tools to shape our future selves warrants thorough investigation and calls for a commitment to research and design practices that prioritize ethical considerations and the long-term benefits of AIMC tools.

### 5.4 Limitations and Future Work

We recognize several limitations in our study. First, although some participants used integrated writing companions like the Grammarly desktop app, for many participants, who had to open a new browser window to access AIMC services like ChatGPT and Quillbot, their user experience might have been compromised. Second, the study relies on self-reported data, raising concerns about selection bias. Participants may have selectively reported specific types of communication, possibly omitting informal or brief exchanges, thereby skewing the dataset. Third, the duration of our diary study was limited to one week, and participants would almost certainly learn more about the potential long-term effects of using AIMC tools had the study continued over a longer period of time. Fourth, although we collected 274 daily survey entries from participants, the small number of participants is a limitation of our study. Finally, while participants' commonly used AIMC tools were collected in the screening survey and detailed in Table 3, we did not specifically gather information on the tools used by participants during the diary study. We noted from the exit interview that most of our participants used ChatGPT (both paid and unpaid versions), but we did not have a specific question probing which tool they used for each diary entry. We acknowledge that the majority use of ChatGPT may bias our results, and different AI models may behave differently.

Future research could employ in-situ data collection methods to capture a broader range of user communications. Additionally, co-design sessions with users to develop AIMC tools that better align with their specific goals and needs would be a valuable complement to the data presented here.

## 6 CONCLUSION

Using mixed methods, we conducted a one-week diary and interview study to understand users' experiences using AIMC tools for interpersonal communication. Our research reveals that users generally appreciate AIMC assistance, noting advantages such as increased confidence and support for finding the words to express themselves with precision. Participants used these tools to generate communication ideas, understand cross-cultural communication differences, and interpret the emotions behind the messages they received from others.

Yet, our study also highlights drawbacks of existing AIMC tools, such as their tendency to generate verbose, unnatural responses and overly emotional content. These problems amplify users' concern about the authenticity of AIMC-generated messages, and the potential for users to become overly dependent on this technology over time.

Additionally, we pinpointed four critical communication spaces where users' perceptions of AIMC tools diverge, categorized by the level of communication stakes (high or low) and relationship dynamics (formal or informal). Participants found these tools more suitable for formal interactions than casual ones. They also noted that this support is particularly valuable in high-stakes situations, where they benefit from the tool's coaching and content suggestions.

Overall, our study provides insights into how users perceive the role of AIMC tools in interpersonal communication. We explore users' reflections on both the short- and long-term impacts of these tools and provide recommendations for designers and researchers of these systems.

## ACKNOWLEDGMENTS

Thanks to Yifan Lin, Lucia Fang, and Lynn Nguyen for their help with this research. This work was funded by Jacobs Foundation Research Fellowship to Alexis Hiniker.

*

# A  APPENDIX A: PARTICIPANT TABLE

Table 3: Participant demographics, fluency in English, and experience of using AI for interpersonal expression.

| PID | Age Group | Gender | Familiar AIMC Tools | Non-native Speaker | Frequency of AIMC for Interpersonal Communication | Highest Education |
| --- | --- | --- | --- | --- | --- | --- |
| P1 | 18-20 | Female | ChatGPT, Dalle-E | No | Several times /month | Some college |
| P2 | 21-29 | Prefer not to say | ChatGPT | Yes | Don't use | Master's degree |
| P3 | 21-29 | Female | Chat GPT, Canva AI, Beautiful AI | Yes | Several times /week | Bachelor's degree |
| P4 | 21-29 | Female | ChatGPT, NotionAi | No | Several times /week | Master's degree |
| P5 | 21-29 | Female | ChatGPT | No | Several times /month | Bachelor's degree |
| P6 | 30-39 | Prefer not to say | chatGPT | Yes | less than once a month | Master's degree |
| P7 | 21-29 | Male | Quilbot, ChatGPT, NotionAI | Yes | Several times /week | Bachelor's degree |
| P8 | 21-29 | Female | ChatGPT | No | Several times /week | Bachelor's degree |
| P9 | 21-29 | Female | ChatGPT, Grammarly, GrammarlyGO, NotionAI | Yes | At least once a day | Master's degree |
| P10 | 18-20 | Female | ChatGPT | No | Don't use | Some college |
| P11 | 30-39 | Male | Chat GPT, Gooey.AI | Yes but use English primarily | At least once a day | Master's degree |
| P12 | 21-29 | Female | ChatGPT, Grammarly | Yes | At least once a day | Master's degree |
| P13 | 21-29 | Female | GrammarlyGO > ChatGPT > NotionAI | Yes | Several times /week | Master's degree |
| P14 | 21-29 | Female | ChatGPT, Canva AI, ResumAI, Quillbot | No | Several times /week | Master's degree |
| P15 | 30-39 | Male | Chat GPT, Notion | Yes | Several times /week | Master's degree |

(a)

(b)

Figure 6: Part of one design prototype (prototype one) we showed to participants. It shows an AIMC tools' setup page that users can customize their specific settings.